\documentstyle[preprint,prb,aps]{revtex}

\begin{document}                             

\title{
  Evidence for Kondo Effect in Au\(_{80}\)Co\(_{20}\) Ribbons}

\draft

\author{
  D.S.~Geoghegan,\footnote{If you have read this paper and wish to be 
  included in a mailing list (or other means of communication) that I 
  maintain on the subject, then send e-mail to: seang@ifw-dresden.de}
  A.~H\"utten,
  K.--H.~M\"uller,
  and L.~Schultz}
\address{
  IFW Dresden, Helmholtzstra{\ss}e 20, D--01069 Dresden, Germany}

\date{3 March 1997}

\maketitle


\begin{abstract}

A minimum in resistivity as a function of temperature for an 
as--quenched Au\(_{80}\)Co\(_{20}\) ribbon prepared by melt--spinning 
using a wheel surface speed of 20~m~s\(^{-1}\) is found at 25~K.  No 
resistivity minimum is found for an as--quenched ribbon using a wheel 
surface speed of 60~m~s\(^{-1}\), however, upon heat treatment of this 
ribbon a resistivity minimum is recovered.  The temperature of the 
minimum decreases with increasing total time of heat treatment.  These 
observations are interpretted as evidence for the microstructural 
control of the Kondo effect typically found in dilute magnetic alloys 
in a giant magnetoresistance granular material.

\end{abstract}



Giant magnetoresistance (GMR) is one mechanism controlling the 
magnetic field dependence of resistivity, \(\rho\), of a material in 
which spin dependent scattering occurs within the bulk of, and at 
interfaces separating, non--magnetizable (NM) and magnetizable (M) 
regions.  GMR was first observed in a multilayer (ML) of Fe/Cr 
\cite{Baibi88a} and later in granular materials. 
\cite{Berko92a,Xiao92a}  Among the models used to describe GMR and 
magnetization in these materials are models based on 
superparamagnetism (SPM). \cite{Zhang93a,Hicke96a,Lucin96a,Wiser96a}  
The SPM models do not adequately describe the low temperature and high 
field behaviour of ultra--fine M entities dispersed in a conducting NM 
matrix.  To explain this descrepancy, an almost two orders of 
magnitude increase in the magnetocrystalline anisotropy energy density 
has been used \cite{Hicke96a} although equally possible is that the 
average magnetic moment per M ion is reduced, as has been observed in 
Au--Co granular GMR ribbons. \cite{Hutte95b}

The Kondo effect \cite{Kondo69a} results in a reduction of the 
effective magnetic moment of individual M ions due to localization of 
conduction electrons of opposite spin.  It has been reported that 
evidence of Kondo scattering at low temperatures can be found in 
Au--Fe GMR materials. \cite{Sato93a,Giord96a}  Since most NM--M 
alloys used at the basis for GMR materials exhibit Kondo scattering in 
the dilute limit \cite{Rossi91a} and local concentrations of M ions in 
the NM matrix may be below the concentration limit for the Kondo 
effect to occur (M ion concentrations typically \(<0.1\)~at\%), the 
expectation of a reduction in average M ion magnetic moment in GMR 
materials is not unreasonable.  There are three identifiers of the 
Kondo effect: (i) a minimum in \(\rho\) below the Kondo temperature, 
\(T_{K}\), below which \(\rho \propto -\ln(T)\), (ii) a specific heat 
singularity at \(T_{K}\), and (iii) a reduction in the M ion magnetic 
moment identifiable through magnetic susceptibility measurements in 
which the Curie--Weiss law is not followed.

The Au--Co binary alloy shows eutectic decomposition at 
Au\(_{75.1}\)Co\(_{24.9}\) and \(T=996.5^{\circ}\)C.  H\"utten 
\emph{et al.} \cite{Hutte95b} estimated the upper limit of the 
chemical and coherent spinodols for Au--Co.  Using these calculations, 
alloys of Au\(_{80}\)Co\(_{20}\) should decompose via a eutectic 
decomposition rather than following a spinodal mechanism when heat 
treated at \(400^{\circ}\)C.  The solubility of Co in Au below 
\(422^{\circ}\)C is very limited and much smaller than 0.2~at\% 
\cite{Okamo85a} hence a Au--Co alloy heat treated at \(400^{\circ}\)C 
should decompose into essentially pure Au and Co.  Microstructural 
studies of rapidly quenched Au--Co alloys 
\cite{Hutte95b,Berna95a,Katao95a} have found that, for Co 
concentraions greater than 15~at\%, Au--rich grains are found in 
the as--quenched state, the boundaries of which are decorated with 
heterogeneously formed spherical fcc essentially pure Co precipitates 
with diameters of up to 100~nm.  The grain interiors are characterised 
by a lamellae eutectic consisting of Co--rich lamellae separated by 
Au--rich lamellae.  Annealing at \(480^{\circ}\)C disolves the 
eutectic and large essentially pure Co precipitates are formed at the 
grain boundaries and within the grains through the transfer of Co ions 
from the Au--rich lamellae to the Co--rich lamellae and finally to the 
Co precipitates using the Co--rich lamellae as diffusion paths. 
\cite{Berna95a}

Two ingots of Au\(_{80}\)Co\(_{20}\) were prepared by arc melting 
elemental components of greater than 99.9\% purity in a high purity Ar 
atmosphere of 20~kPa to yield a total mass of 2.5 to 3~g per ingot.  
The weight loss during arc melting was less than 2\%.  Melt--spinning 
was performed by ejection of the molten alloy through a 0.6~mm 
diameter round hole perforating the bottom of a quartz crucible, with 
an ejection pressure of 20~kPa and 60~kPa, onto a rotating lightly 
sanded Cu wheel, with a wheel surface speed, \(v_{s}\), of 
20~m~s\(^{-1}\) and 60~m~s\(^{-1}\), respectively.  A chamber pressure 
of less than \(10^{-3}\)~Pa was established prior to ejection of the 
melt and the height of the crucible above the surface of the Cu wheel 
was \(0.40\pm 0.05\)~mm.  Typical ribbon pieces had lengths of 2 to 
30~cm, widths of 0.5 to 1~mm and thicknesses of 20 to 40~\(\mu\)m.

Heat treament of the ribbon quenched using \(v_{s}=60\)~m~s\(^{-1}\) 
was conducted at \(400^{\circ}\)C in a closed quartz tube containing 
10~kPa high purity Ar with the samples contained in Ta boats.  The 
total time of heat treatment, \(t_{ann}\), ranged from 10~min to 4~h 
with quenching to room temperature by immersion of the annealing tube 
in water at intermediate times to retrieve samples at different stages 
of heat treatment.  The composition of all samples was determined by 
microprobe analysis which found the samples to have a composition 
within \(\pm 0.2\)~at\% of the nominal composition.

Measurements of longitudinal MR (L--MR, made with the external field, 
\(H_{ext}\), parallel to the drive current, \(i\)) were made at 
temperatures from 10~K to 300~K in magnetic fields of up to 5~T in a 
Lakeshore 7225 susceptometer with the resistance option and of L--MR 
and transverse MR (T--MR, made with \(H_{ext}\) perpendicular to 
\(i\)) in magnetic fields of up to 7~T in a split coil solenoid from 
Oxford Instruments.  The resistance of the samples was measured by 
the four point contact method from which the MR ratio and \(\rho\) 
were calculated.  Measurements of magnetization were performed using a 
SQUID magnetometer in magnetic fields of up to 4~T from 5~K to 300~K.  
Total AC susceptibility, \(\chi_{tot}\) was measured in the Lakeshore 
7225 susceptometer used for the resistivity measurements with an AC 
field magnitude of 2.5~mT at a frequency of 133~Hz.  Differential 
scanning calorimetry (DSC) measurements were performed from 93~K to 
673~K on a TA Instruments DSC 2910.

The MR ratio, \(\Delta\rho/\rho_{0}\), is defined in this work as
\(\Delta\rho/\rho_{0}=[\rho(H_{ext})-\rho(0)]/\rho(0)\) where 
\(\rho(H_{ext})\) is the resistivity for an external field equal to 
\(H_{ext}\).  \(\rho(H_{ext})\) is measured at a series of \(H_{ext}\) 
values corresponding to a full hysteresis loop (i.e. measurements from 
\(H_{ext}=0 \rightarrow H_{max} \rightarrow -H_{max} \rightarrow 0\) 
where \(H_{max}\) is the maximum external field possible with the 
measurement device) after exposing the samples to the maximum reverse 
field.

The temperature dependence of \(\chi_{tot}\) of the as--quenched 
samples shows a small peak at about 10~K superimposed on a generally 
constant larger value and the field cooled (FC) magnetization curve 
separates from the zero field cooled (ZFC) magnetization curve just 
below room temperature.  These results are consistent with spin--glass 
(SG) behaviour (as was concluded by Kataoka \textit{et al.} 
\cite{Katao95a} for an aged Au\(_{85}\)Co\(_{15}\) foil) as well as 
SPM.  A distinction between either is difficult to establish without a 
knowledge of the microstructure of the Co in the Au--rich lamellae and 
Au--matrix grains.  Although \(\chi_{tot}\) does not follow 
Curie--Weiss behaviour, it is difficult to attribute this to the Kondo 
effect.  The observation of both a peak in \(\chi_{tot}\) and 
splitting of the FC and ZFC magnetization curves, indicating the 
presence of either or both of a SG and SPM particles, is not 
surprising given the complex nature of the microstructure.  DSC 
measurements do not show any specific heat singularity that can be 
associated with a Kondo temperature.

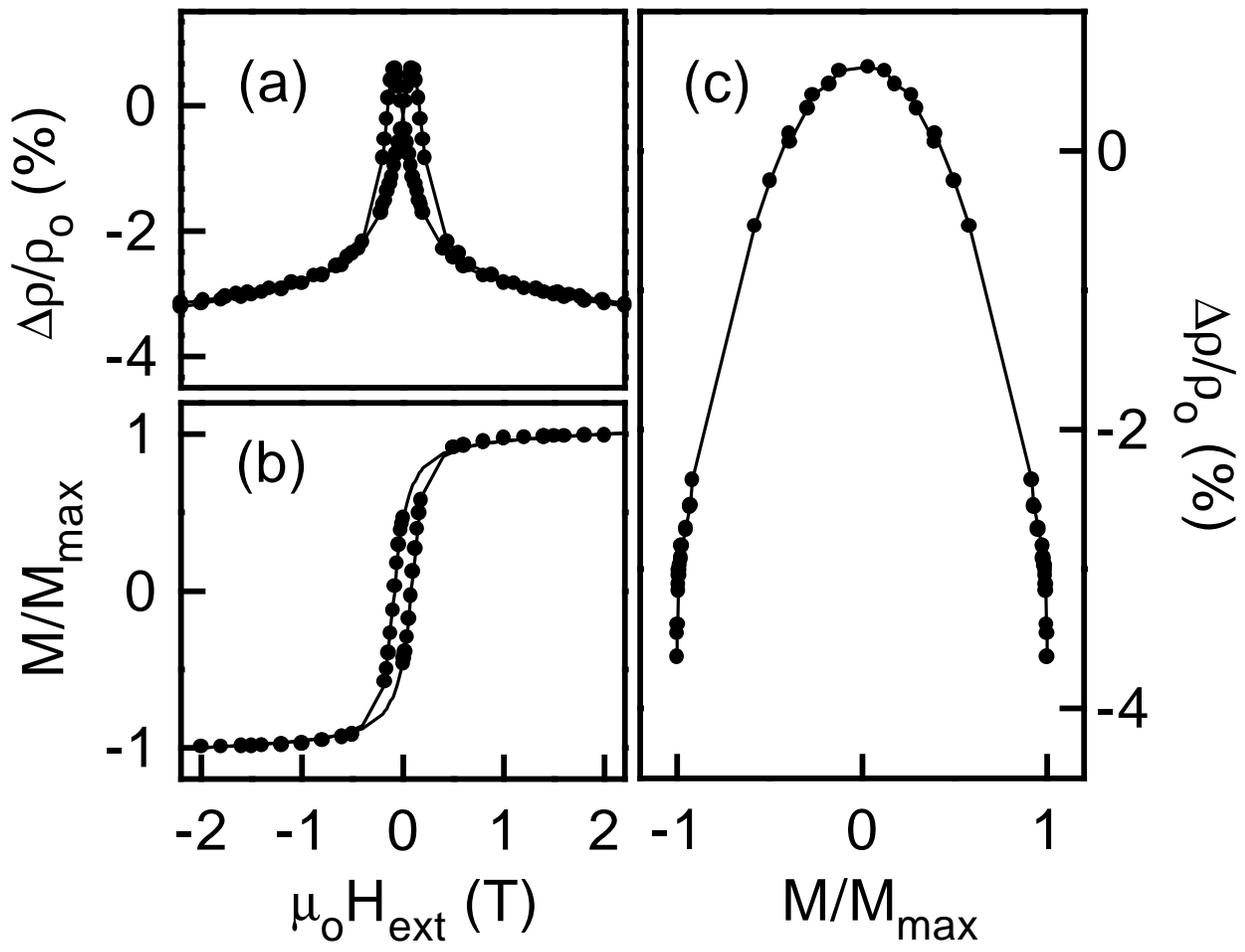
\begin{figure}
  \caption{(a) L--MR, expressed as a MR ratio, as a function of 
  $\mu_{0}H_{ext}$, (b) reduced magnetization, $M/M_{max}$, as a 
  function of $\mu_{0}H_{ext}$, and (c) L--MR as a funtion 
  of $M/M_{max}$, for Au$_{80}$Co$_{20}$ ribbon, melt--spun using 
  $v_{s}=60$~ms$^{-1}$ and heat treated at $400^{\circ}$C for a total 
  time of $40$~min, measured at $T=10$~K.}
  \label{fig:01}
\end{figure}

In Fig.~\ref{fig:01} is shown L--MR, expressed as a MR ratio, as 
a function of \(\mu_{0}H_{ext}\), reduced magnetization, 
\(M/M_{max}\), as a function of \(\mu_{0}H_{ext}\), and L--MR as a 
funtion of \(M/M_{max}\), all measured at \(10\)~K, for 
Au\(_{80}\)Co\(_{20}\) ribbon (\(v_{s}=60\)~m~s\(^{-1}\)) heat treated 
at \(400^{\circ}\)C for a total time of 40~min.  \(M_{max}\) measured 
at 4~T is 9.94~A~m\(^{2}\)/kg which is 82\% of the value expected 
assuming simple dilution of magnetization by the Au, similar to the 
result obtained by H\"utten \emph{et al.} \cite{Hutte95b}  Measurement 
of T--MR shows significant anisotropic MR (AMR) which is due to 
ferromagnetic--like regions in the ribbon, the full results of these 
will be presented elsewhere. \cite{Geogh97a}  The low field GMR is 
ML--like (i.e. proportional to the square of magnetization and 
saturating in low fields) and at high fields the magnetization and the 
GMR increases almost linearly with \(H_{ext}\) consistent with the 
arguments of Wiser \cite{Wiser96a} concerning scattering between SPM 
and blocked M entities in the material.

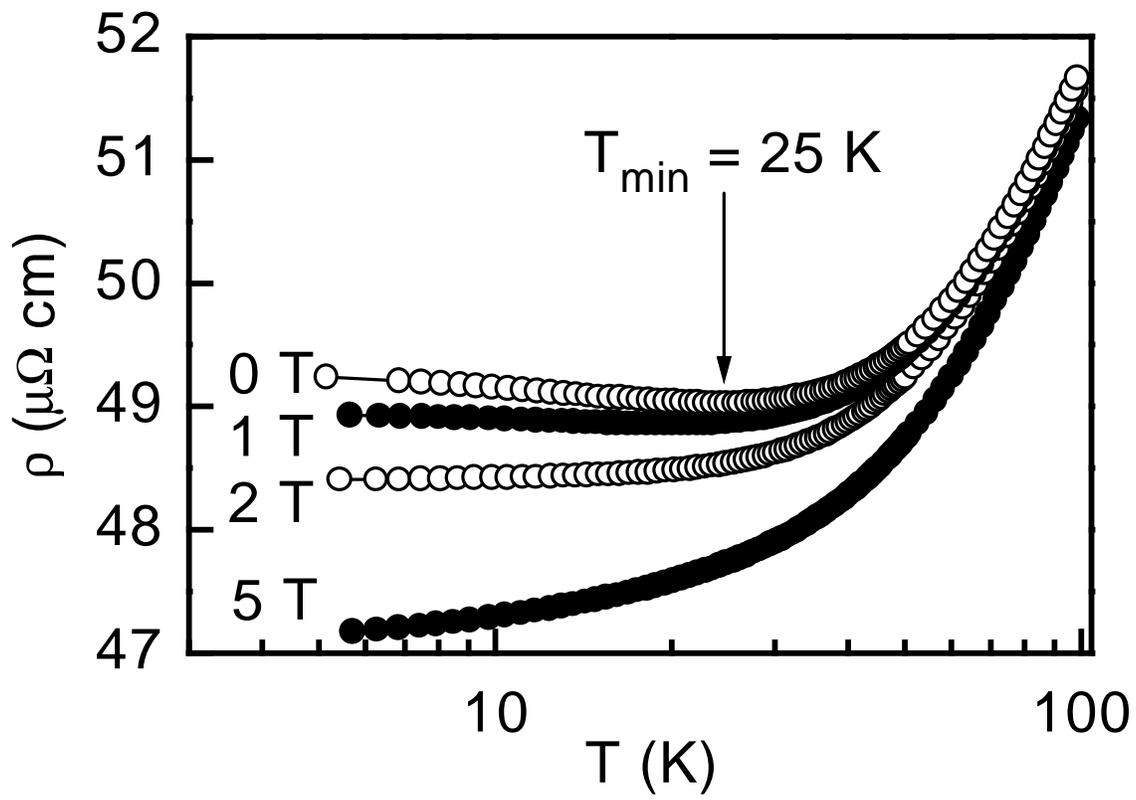
\begin{figure}
  \caption{Resistivity, $\rho$, as a function of temperature for 
  as--quenched Au$_{80}$Co$_{20}$ ribbon, melt--spun using 
  $v_{s}=20$~ms$^{-1}$, in constant external magnetic fields.  
  A minimum in $\rho$ at $T=25$~K and below this minimum 
  $\rho \propto -\ln(T)$ for $H_{ext}=0$.}
  \label{fig:02}
\end{figure}

Shown in Fig.~\ref{fig:02} is \(\rho\) of as--quenched 
Au\(_{80}\)Co\(_{20}\) ribbon (\(v_{s}=20\)~m~s\(^{-1}\)) as a 
function of temperature for constant external magnetic fields.  The 
linear increase in \(\rho\) with the decrease in the logarithm of 
temperature below \(T=25\)~K for \(H_{ext}=0\) suggests that at least 
some of the Co is diluted enough in the Au--rich lamellae and/or 
Au--matrix grains for the Kondo effect to occur.  Kondo scattering at 
these temperatures results in an extra magnetic contribution to 
\(\rho\) and an enhancement of the measured GMR response as can be 
seen from the splitting of the \(\rho\) curves measured at the 
different values of external magnetic field.  Any SG regions would 
also contribute towards the enhancment in GMR at these temperatures.

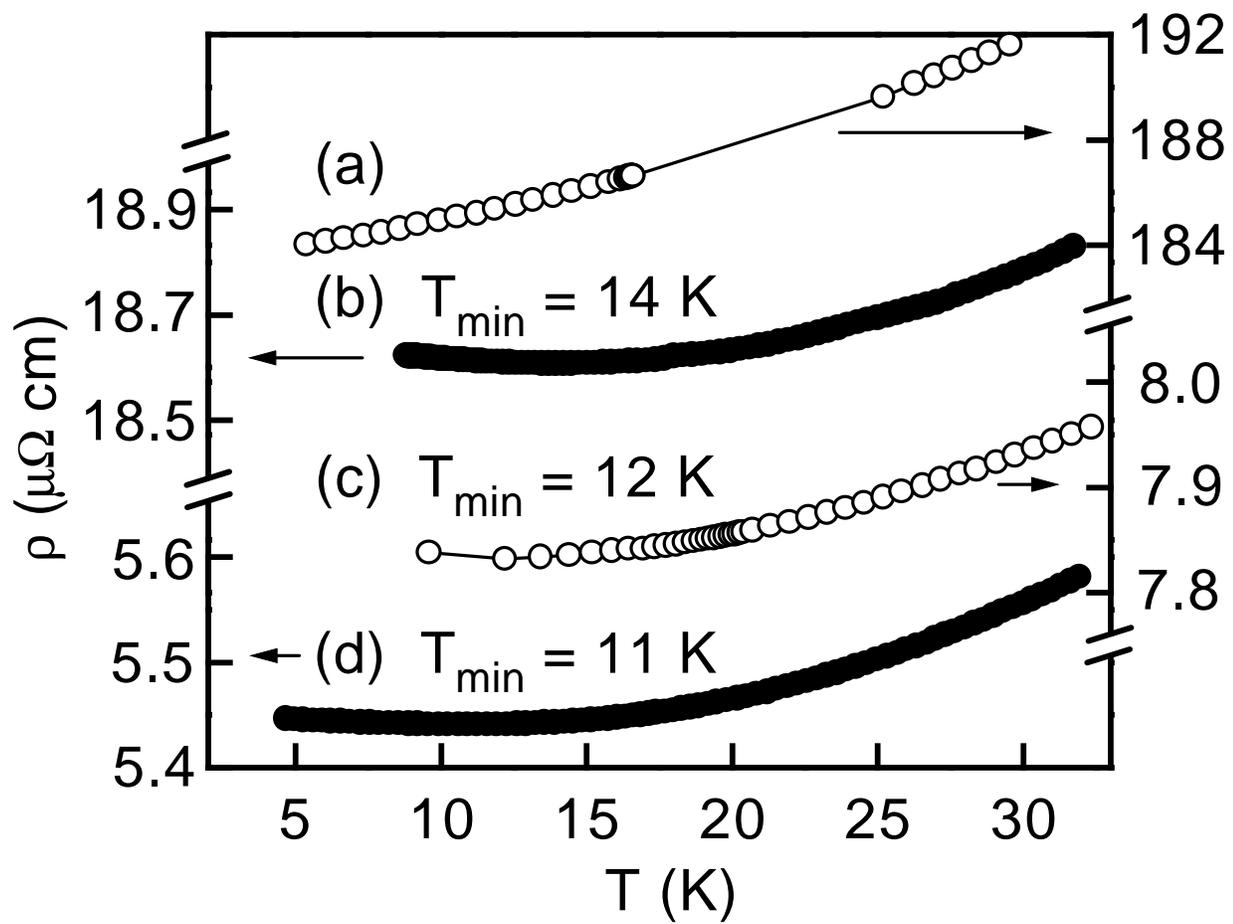
\begin{figure}
  \caption{Resistivity, $\rho$, as a function of temperature for
  Au$_{80}$Co$_{20}$ ribbon, melt--spun using 
  $v_{s}=60$~ms$^{-1}$ and heat treated at $400^{\circ}$C for a total 
  time of: (a) $0$~min (as--quenched), (b) $10$~min, (c) $40$~min, and 
  (d) $4$~h.  Resistivity minima are only found for heat treated 
  ribbons, which occur at: (b) $14$~K, (c) $12$~K, and (d) $11$~K.  
  Note the change of scale for each curve.}
  \label{fig:03}
\end{figure}

In Fig.~\ref{fig:03} is shown the development of \(\rho\) as a 
function of temperature for Au\(_{80}\)Co\(_{20}\) ribbon 
(\(v_{s}=60\)~m~s\(^{-1}\)) heat treated at \(400^{\circ}\)C for a 
total time of: (a) 0~min (as--quenched), (b) 10~min, (c) 40~min, and 
(d) 4~h.  In addition to the general decrease in \(\rho\) with heat 
treatment, minima in \(\rho\) can be found for the heat treated 
ribbons, the temperature of which decreases with increasing 
\(t_{ann}\).  The development of the temperature dependence of 
\(\rho\) with \(t_{ann}\) is not consistent with an increase of a 
contribution from the Kondo effect since: (i) one would naturally 
assume that a decrease of the Co concentration in the Au--rich 
lamellae and/or Au--matrix grains from a increase in \(t_{ann}\) would 
result in an increase in the effective Kondo temperature, and (ii) 
these results are consistent with the interpretation that the minima 
in \(\rho\) are associated with SG transition temperatures since the 
temperature of the SG transition should decrease with decreasing Co 
concentration in the Au--rich lamellae and/or Au--matrix grains.  
However the increase in \(\rho\) with the logarithm of temperature as 
the temperature decreases as shown in Fig.~\ref{fig:02} as well as the 
preceeding minimum in \(\rho\) are indicative of the Kondo effect 
since the M ion magnetic moment correlation length in a SG 
monotonously increases with decreasing temperture until the spin 
freezing temperature is reached where upon it remains almost constant.
  
In conclusion, we have observed in Au\(_{80}\)Co\(_{20}\) ribbons 
prepared by melt--spinning and subsequent heat treatment resistivity 
minima associated with the Kondo effect.  It is possible that the 
reduction of the M ion magnetic moment in GMR materials due to the 
Kondo effect can explain the difficulty in fitting the observed 
magnetic properties to the proposed SPM models and avoid the almost 
two orders of magnitude increase in magnetocrystalline anisotropy 
energy density \cite{Hicke96a} particularly if the Kondo effect 
results in significant reduction in the effective magnetic moment of 
the M ions contributing to the magnetic and resitivitiy changes at 
high fields in these materials.  Determination of the Co 
concentrations in the Au--rich lamellae would confirm these 
conclusions which is expected to be accomplished soon.

We would like to thank Dr.~B.~Idziwokski for discussions, 
Dr.~D.~Eckert for performing the magnetization measurements, and 
H.~Schuffenhauer for performing the microprobe analysis.
 


\end{document}